\documentclass[%
prl%
,twocolumngrid%
,secnumarabic%
,amssymb]{revtex4}

\usepackage{amsmath}%
\usepackage{bm}%

\usepackage[T1]{fontenc}
\usepackage{graphicx}
\newcommand{\tab}{\makebox[4em]{}}

\begin{document}

\preprint{}

\title[Short Title]{Coexistence of Global and Local Time 
Provides Two Ages for the Universe}

\author{Gary D. Warren}

\email{gary.warren@saic.com}
\affiliation{Science Applications International Corporation}

\date{\today}

\begin{abstract}
Lorentz time is replaced by three parameters: a global time, a local rate of aging, and a new spatial 
coordinate. This enables successful reconstruction of relativity's observables in a four dimensional 
Euclidean hyperspace. The new formulation predicts that the age of the universe, as inferred from the 
Hubble Constant, is less than the observed evolutionary age of astronomical configurations, as has 
been repeatedly measured. It also predicts a new phenomenon that provides closure of the universe 
without introduction of unobserved dark matter and independent of gravity. Further, it predicts 
observation of an accelerating expansion rate of the universe. It also explains our ability to perceive 
only three of the four spatial coordinates and predicts parallel, mutually unobserved universes that 
evolve independently while being correlated as past, present, and future.
\end{abstract}

\pacs{Valid PACS appear here}%

\maketitle
\tableofcontents

\section{The Coordinate System}

In space are observers about which we know the following experimentally: Each observer observes three 
locally flat spatial coordinates \ensuremath{(x, y, z)} and a local rate of aging, \ensuremath{dt}. Each clock 
displays a local time, \ensuremath{t}, that is the integral of \ensuremath{dt} 
as experienced by that clock. The time shown on each clock, also known as local 
time, is the amount of system evolution experienced by the local observer, and the rate of system 
evolution varies from clock to clock as known from relativity.

Special relativity transmutes the time shown on clocks into a physical dimension. The local \ensuremath{t}
is merged with local spatial coordinates \ensuremath{(x,y,z)} to create an \ensuremath{(x,y,z,t)} space-time, 
and the principle of relativity is used to determine the covariance of space and time in coordinate 
transformations.  This results in the Lorentz metric. It also results in the absence of global 
time and, as a consequence, creates the inability to embed the universe in a global \ensuremath{(W,X,Y,Z,T)} 
Euclidean hyperspace.\cite{Gravitation}

The approach here does not merge space and time; it leaves local time, \ensuremath{t}, as simply the 
integral of \ensuremath{dt}, which we know it to be. \ensuremath{dt} is a scalar function of location
corresponding to the rate of system evolution at each location. That is, \ensuremath{dt} is the local rate of
aging. 

A new global spatial coordinate, \ensuremath{W}, separate from time, is added to three global \ensuremath{(X,Y,Z)} Euclidean coordinates to create a global \ensuremath{(W,X,Y,Z)} Euclidean hyperspace. The four coordinates obey Euclidean translation and rotation invariances. The space may also referenced in four-dimensional spherical or other coordinates without loss of generality, as in Section{\nobreakspace}3. Whereas in Kaluza-Klein theories one of the spatial dimensions is compacted\cite{KK}, in this theory each spatial coordinate \ensuremath{(W, X, Y, Z)} has infinite extent. The metric is:
\begin{equation}
\label{eq:metric}
ds^{2}=dW^{2}+dX^{2}+dY^{2}+dZ^{2}{\nobreakspace}.
\end{equation}
Time is not a part of the metric. All points in the space are considered physically real and take on only real values. The observed universe is embedded in the Euclidean space as a quasi-three dimensional surface \ensuremath{F_{0}(W,X,Y,Z)< \delta}. The curvature of the observable universe is assumed to be smooth (differentiable) with its thickness, \ensuremath{\delta}, small relative to the curvature of the surface. This is simply a generalization of the observation that space is everywhere locally observed as flat. 

One global time, \ensuremath{T}, is defined such that all events can be uniquely referenced by 
\ensuremath{(W,X,Y,Z,T)}. \ensuremath{T} is not, however, a spatial coordinate. That is, events 
that happen at time \ensuremath{T} at location \ensuremath{(W,X,Y,Z)}, happen and are gone. They do not 
continue to reside at a point \ensuremath{(W,X,Y,Z,T)} to which we can travel. The global time, 
\ensuremath{T}, defines simultaneity. That is, two events that happen at the 
same time \ensuremath{T} are simultaneous. The fact that different observers may see them as 
happening at different local times and in a different sequence becomes, in this formulation, an 
artifact of the observation process. The relationship between the rate of local time 
and the rate of global time is given by \ensuremath{dt/dT}; and it becomes, in this formulation
a property of the physical system rather than a property of the coordinate system. (See Section 2.)

The observable universe, \ensuremath{F_{0}}, has motion in accord with kinematic rules. It is one of an infinite set of locally parallel observable universes \{\ensuremath{F_{i}, -\infty < i < \infty}\} that fill the Euclidean space. We observe only the one sub-space \ensuremath{F_{0}} for physical reasons determined in Section 3.2. Other observers may observe other \ensuremath{F_{i}}.  Local coordinates \ensuremath{(x,y,z)}, defined 
within each \ensuremath{F_{i}}, are curved to follow the shape of \ensuremath{F_{i}} and uniquely reference each location in \ensuremath{F_{i}}. Where \ensuremath{F_{i}} is locally flat, then the local \ensuremath{(x,y,z)} can be overlaid on the global \ensuremath{(X,Y,Z)} making them equivalent within the locally flat region.

\section{Reconstruction of Special Relativity as Observables}

The principle of relativity is used together with the observed kinematics on particles (time dilation, 
particle anti-particle pair formation) to determine
aspects of the kinematics of the observable universe as embedded in the \ensuremath{(W,X,Y,Z)} 
hyperspace. As used here, the principle of relativity requires that no point or direction in space 
be distinguishable from any other by any local measurement. 

\subsection{Observing Particles}

\begin{figure*}
\includegraphics[width=7in]{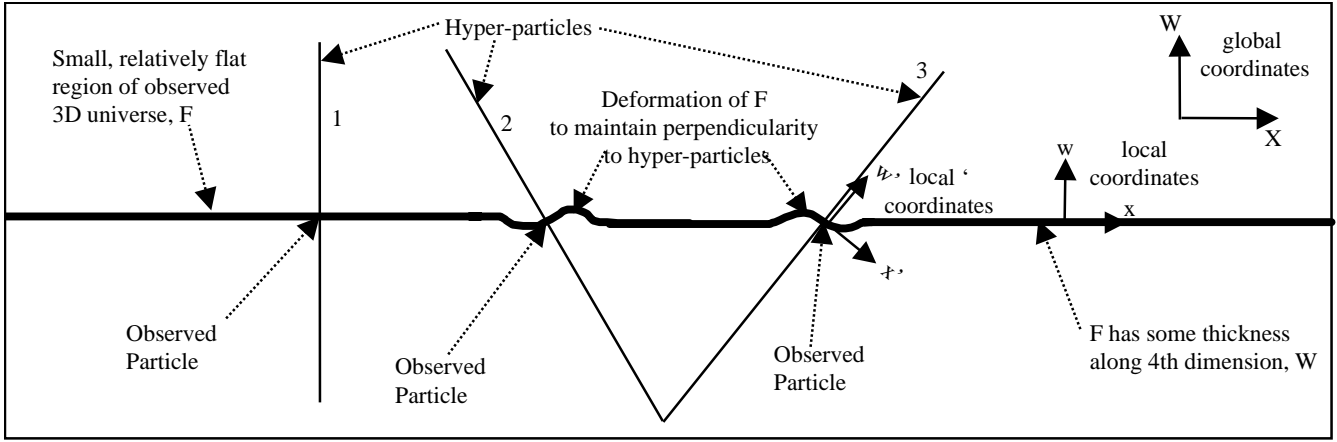}
\caption{A snapshot in time of a two dimensional slice through
the \ensuremath{(W,X,Y,Z)} space. The slice is chosen perpendicular to
the  observable universe, \ensuremath{F_{0}}, showing a quasi one-dimensional
cut through \ensuremath{F_{0}}. The snapshot also shows three hyper-particles
contained in the slice and the particles observed at the intersection
of those hyper-particles with the observable universe. Near hyper-particles 2 and 3, 
the observable universe is deformed to maintain perpendicularity between the 
hyper-particles and the observable universe. The global \ensuremath{X} coordinate is,
for convenience, oriented to align with the local coordinates, away from the 
deformations at hyper-particles 2 and 3.}

\end{figure*}

The observation of particles implies the existence of hyper-particles
in the global space. Hyper-particles are  quasi-one-dimensional
entities of various types such that, at the intersection of a
hyper-particle and the observable universe, we observe a quasi-zero-dimensional particle
corresponding to a known particle type. The hyper-particles
are not assumed stationary; they are not like ``world-lines'' defined
in other formulations\cite{Gravitation}. Rather, the hyper-particles move and obey
kinematics rules to be derived later in the discussion. Figure
1 shows a snapshot in time of hyper-particles intersecting an
observable universe to create observable particles at the intersection.

The principle of relativity requires that hyper-particles be
everywhere perpendicular to the local direction of the observed
universe. If hyper-particles were not everywhere perpendicular
to the local observed universe, then a local measurement could
distinguish one direction in the observable universe from all
others, in violation of relativity. Thus, in Figure{\nobreakspace}1, the observable
universe is shown deforming near each hyper-particle to be locally
perpendicular to it. 

The principle of relativity requires that an observer be able to reside at any point in 
\ensuremath{(W,X,Y,Z)} space. At that point in space, the observer must, further, reside in some \ensuremath{F_{i}}. 
Otherwise some locations in space would be distinctly different from others, in violation
of relativity. Thus, the global space is filled with observable
universes. This implies that each observable universe has a four dimensional
volume. Thus, each observable universe is described as a three
dimensional surface plus some thickness, \ensuremath{\epsilon} (related to \ensuremath{\delta}), along a
local \ensuremath{w} coordinate locally perpendicular to that surface, as
shown in Figure{\nobreakspace}2a. Then an infinite set of parallel universes
fill the \ensuremath{(W,X,Y,Z)} space as shown in Figure{\nobreakspace}2b.

\begin{figure*}
\includegraphics[width=7in]{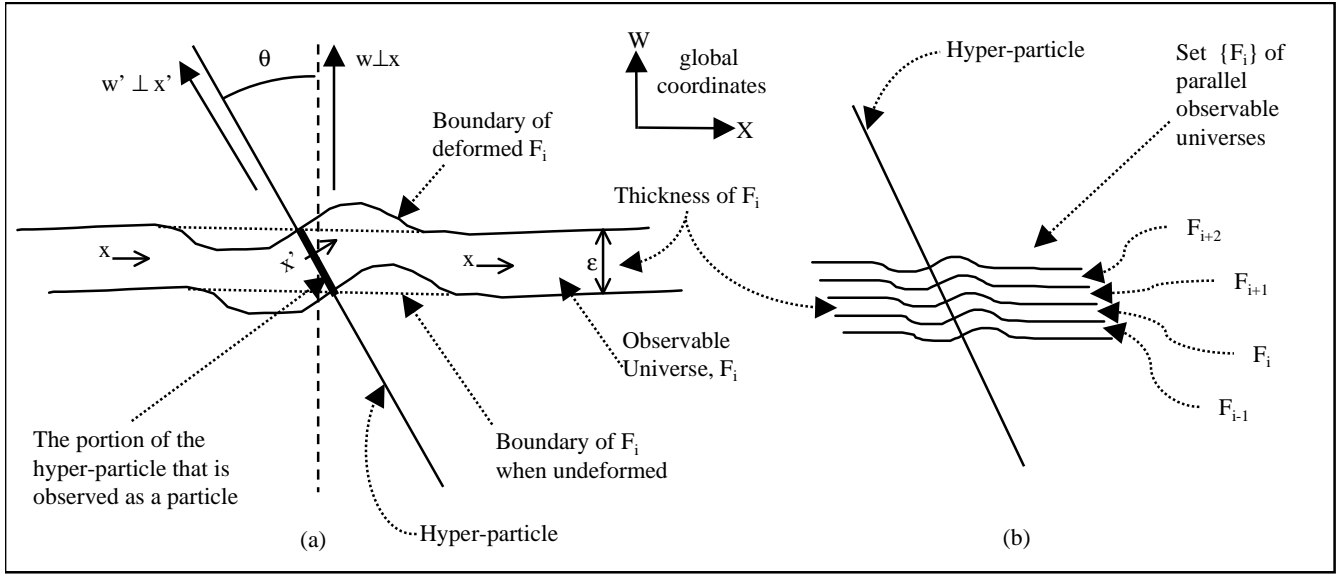}
\caption{The intersection of a hyper-particle with observable
universes. (a) Shows the deformation of \ensuremath{F_{i}} and the corresponding rotation of the local
coordinate \ensuremath{w} to \ensuremath{w'}. (b) Shows a set of 
adjacent parallel observable universes deforming near a hyper-particle, while maintaining identical shapes 
and without forming gaps between them.}
\end{figure*}

The principle of relativity requires that the thickness of observable
universes increase at intersections with hyper-particles not perpendicular
to the overall direction of the observable universe, as in Figure{\nobreakspace}2. 
If the thickness did not increase, then either (1) local gaps would form between 
adjacent observable universes, creating locations not within an observable
universe, in violation of relativity, or (2) the shapes of \ensuremath{F_{i-1}}, 
\ensuremath{F_{i}}, and \ensuremath{F_{i+1}} would have to be different
from each other in ways that make them distinguishable, in violation of relativity. 
By inspection of Figure{\nobreakspace}2a, a thickness of \ensuremath{\epsilon/cos\theta} for 
\ensuremath{F_{i}} provides accordance with relativity. Here \ensuremath{\theta} is the angle 
of deformation caused by the hyper-particle.  
  
\subsection{Observing Time Dilation}

By relativity, the rate of local time at any location must vary inversely as the local thickness 
of the observable universe. Otherwise a local measurement of the thickness of the universe
using, for instance, the time it takes for light to cross the local thickness of that observable 
universe, would distinguish one point in space from another by local measurement, which would
violate relativity. Thus,
\begin{equation}
\ensuremath{\epsilon}\ensuremath{\chi}/c = constant  \tab ,
\label{eq:1}
\end{equation}
where, \ensuremath{\chi \equiv dt/dT} is the local ratio of the rate
of local time to the rate of global time, \ensuremath{dT} is a global constant,
\ensuremath{c} is the local speed of light; and \ensuremath{\epsilon} is the local thickness
of the observable universe at any particular location in space.
In Equation{\nobreakspace}(\ref{eq:1}), \ensuremath{c} and \ensuremath{\epsilon} are both measured in global coordinates.
From Equation{\nobreakspace}(\ref{eq:1}), it follows that the rate of local time is invariant
under Euclidean transformations. This is because the rate of
local time depends on the scalar quantities, \ensuremath{c} and \ensuremath{\epsilon}, that
are invariant under Euclidean spatial transformation. It also
follows that the observable universe must also stretch along
the \ensuremath{x}, \ensuremath{y}, and \ensuremath{z} coordinates 
by a factor \ensuremath{1/\chi} so that the
observed speed of light remains everywhere the same.

By Equation{\nobreakspace}(\ref{eq:1}), the local rate of time in the region near the
hyper-particle in Figure{\nobreakspace}2a, relative to the local rate of time
away from the hyper-particle, is given by
\begin{equation}
 dt' = dt cos\theta		\tab	  .
\label{eq:2}
\end{equation}

The local slowing of the rate of time is known, in special relativity,
as ``time dilation''. Specifically, time dilation in Lorentz coordinates
gives a change in the local rate of time as \ensuremath{dt'=dt/\gamma},
where \ensuremath{\gamma=(1-v^{2}/c^{2})^{-1/2}}. 
Comparing, time dilation for special relativity with Equation{\nobreakspace}(\ref{eq:2}) gives
\begin{equation}
\label{eq:3}
dt'/dt = 1/\gamma = cos\theta \tab ,
\end{equation}
which implies that the observed velocity for any particle relates
to the angle of that hyper-particle as
\begin{equation}
\label{eq:4}
v = \pm c sin\theta \tab .
\end{equation}

Thus, the larger the angle, \ensuremath{\theta}, of the hyper-particle,
the faster the observed speed of the observed particle. Equation{\nobreakspace}(\ref{eq:4}) includes an ambiguity regarding whether, for a given angle,
the observed particle travels towards positive or towards negative
\ensuremath{x}.

\subsection{Observing Finite Lifetime}

An observed particle may have finite lifetime due to a number of causes.
One cause can be that a particle is observed to come into existence
in any particular observable universe when a hyper-particle first
intersects that particular observable universe, and is destroyed
when the hyper-particle no longer intersects that observable
universe. The lifetime, \ensuremath{\delta t}, of such a particle is given by
\begin{equation}
\label{eq:5}
\delta t = L/ V_{T} \tab ,
\end{equation}
where \ensuremath{L} is the length of the hyper-particle and 
\ensuremath{V_{T}} is its velocity
 through the observable universe (measured in global coordinates). By
relativity, \ensuremath{\delta}t, as measured by a clock at the particle, must
be the same for all particles. Otherwise one point in space would
be distinguishable from another by local measurement. Similarly, by relativity, \ensuremath{V_{T}}
must be the same for all hyper-particles. Otherwise a local measurement
of the time it takes a particular point on a hyper-particle to
pass through the observable universe would not give the same
result everywhere. [Even though the thickness of the observable
universe varies as \ensuremath{\epsilon}/cos\ensuremath{\theta}, the local rate of
time varies also as t/cos\ensuremath{\theta}. Hence the velocity of hyper-particle
through the observable universe proportional to \ensuremath{\epsilon/t},
which is independent of \ensuremath{\theta}.] Finally, since the clock
rate varies as per Equation{\nobreakspace}(\ref{eq:1}), thus, to satisfy Equation{\nobreakspace}(\ref{eq:5}) for
all velocities of the particle, the length of the hyper-particle
must be L=\ensuremath{\gamma L_{0}}, where \ensuremath{L_{0}} is the length of the hyper-particle
for that particle when at rest. Thus, the length of the hyper-particle
increases in proportion to \ensuremath{\gamma}, which causes the observed lifetime of the observed particle
to increase in proportion to \ensuremath{\gamma}, in accord with relativity.

\subsection{Observing Particle Anti-Particle Pair Formation}

The observance of particle anti-particle pair formation provides
one determination for \ensuremath{V_{T}}. It also provides one resolution for
the ``\ensuremath{\pm}'' direction ambiguity for the observed
velocity, \ensuremath{v}, in Equation{\nobreakspace}(\ref{eq:4}). Referring to Figure{\nobreakspace}1, 
hyper-particles 2 and 3 touch at a point below the observed universe, \ensuremath{F_{0}}. If the
hyper-particles are moving towards positive \ensuremath{w}, then, when that
point goes through \ensuremath{F_{0}}, the observed effect will be
of two particles colliding and annihilating --- as particle and
anti-particle. If they are moving towards negative \ensuremath{w}, then they
were seen already to come into existence via pair formation,
and are seen as moving away from each other. 

The event of particle anti-particle pair formation will be seen
in each observable universe, \ensuremath{F_{i}}, if hyper-particles 2 and 3 stay
exactly touching each other (as in Figure{\nobreakspace}1) while flowing through \{\ensuremath{F_{i}}\}. 
That is achieved if the hyper-particle velocity along \ensuremath{x} is zero. Thus, the hyper-particle
motion is given by:
\begin{equation}
\label{eq:6}
V_{W} = V_{T}cos\theta \tab , {\nobreakspace}and
\end{equation}
\begin{equation}
\label{eq:7}
V_{X} = -V_{T}sin\theta + v = -V_{T}sin \theta \pm c sin \theta
= 0 {\nobreakspace} .
\end{equation}
Equation{\nobreakspace}(\ref{eq:7}) is true only if 
\begin{equation}
\label{eq:8}
V_{T} = \pm c \tab and \tab v= V_{T}sin \theta {\nobreakspace} . 
\end{equation}
Thus, the hyper-particle passes through the observable universe
at speed c in one direction or another. The ``\ensuremath{\pm}{\textquotedbl}
ambiguity of Equation{\nobreakspace}(\ref{eq:4}) is now seen to imply that hyper-particles
may in general, flow in either direction through the observed universe.

\subsection{Lorentz Metric}
\label{sec:LM}

While the metric for the space is strictly Euclidean as per Equation{\nobreakspace}(\ref{eq:metric}), the Lorentz metric also appears in the theory. However, the Lorentz metric is not the metric of the space, it is rather an observable of the relationship between events in the space. 

Figure{\nobreakspace}3 overlays a snapshot of a hyper-particle segment  at two instants of time, \ensuremath{T} and \ensuremath{T'}. The segment is at \ensuremath{\overline{AD}} at global time, \ensuremath{T}, and at \ensuremath{\overline{BC}} at a global time, \ensuremath{T'}. If \ensuremath{V_{T} = -c}, then the hyperparticle moves from \ensuremath{\overline{AD}} to \ensuremath{\overline{BC}} over the time interval (T,T') and T'>T. If \ensuremath{V_{T} = +c}, then the hyperparticle moves from \ensuremath{\overline{BC}} to \ensuremath{\overline{AD}} over the time interval (T',T) and T>T'.

Moving in accord with Equations{\nobreakspace}(\ref{eq:6}) through (\ref{eq:8}) the hyper-particle motion creates a set of events in the \ensuremath{(W,X,Y,Z)} universe. \ensuremath{A} is the event of observing a particle at location \ensuremath{(W,X,Y,Z)} at global time \ensuremath{T}. \ensuremath{B} is that same event
observed at the same 3D location in a different observable universe at location \ensuremath{(W',X,Y,Z)} at global time \ensuremath{T'}. Event \ensuremath{C} is the event of observing the same particle at location \ensuremath{(W,X',Y,Z)} at time \ensuremath{T'} that is observed in event \ensuremath{A} at location \ensuremath{(W,X,Y,Z)} at \ensuremath{T}. Included in the figure are two different observable universes: (1) the observable universe containing event \ensuremath{A}, in the shape it has at time \ensuremath{T}; and (2) the observable universe containing event \ensuremath{B}, in the shape it has at time \ensuremath{T'}.

\begin{figure*}
\includegraphics[width=7in]{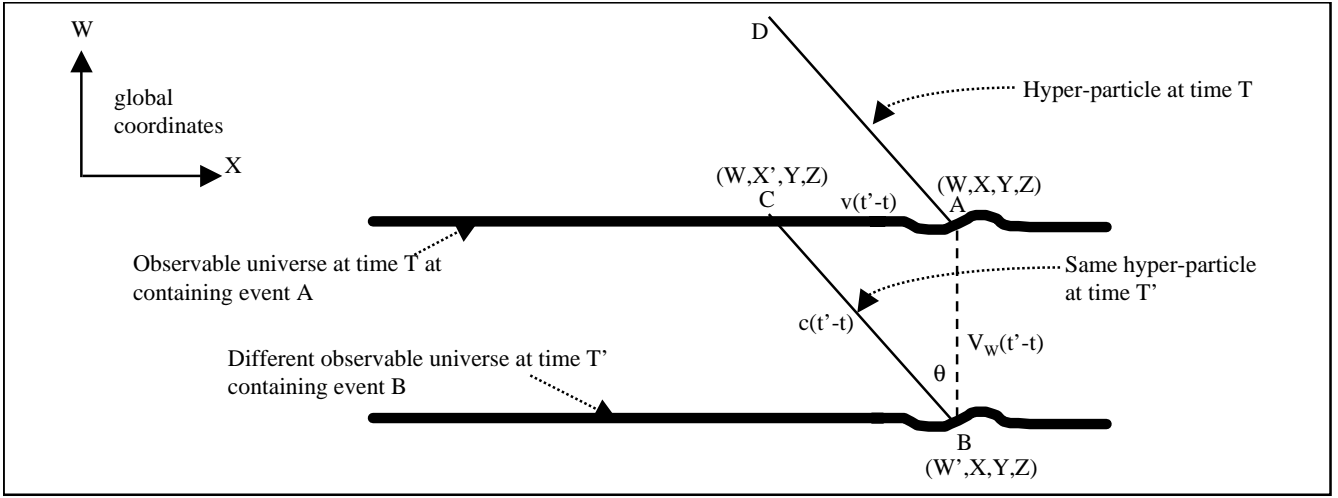}
\caption{Hyper-particle segment \ensuremath{\overline{AD}} moves to \ensuremath{\overline{BC}} over time \ensuremath{(T'-T)}. The Lorentz metric, in this theory, is not the metric of the space, but rather relates events occurring at locations \ensuremath{A}, \ensuremath{B}, and \ensuremath{C}. The metric of the space, in this theory, is strictly Euclidean as per Equation{\nobreakspace}(\ref{eq:metric}).}
\end{figure*}

Setting \ensuremath{dT=\mid T'-T \mid} in Figure{\nobreakspace}3, the distance \ensuremath{\mid \overline{AC}\mid } is the distance which the observed particle travels in time \ensuremath{dT}. That distance 
is \ensuremath{vdT}. The distance \ensuremath{\mid \overline{BC}\mid } is equal to the amount of hyper-particle that has passed through the upper observable universe in time \ensuremath{dT}. That distance is \ensuremath{V_{T}dT = cdT}. The event \ensuremath{A} moves to event \ensuremath{B} at rate \ensuremath{V_{W}} as given by Equation{\nobreakspace}(\ref{eq:6}). Thus, sides of triangle \ensuremath{\overline{ABC}} have 
lengths whose squares are:
\begin{equation}
\label{eq:9}
\mid\overline{BC}\mid^{2} = c^{2} dT^{2}{\nobreakspace},
\end{equation}
\begin{equation}
\label{eq:10}
\mid\overline{AC}\mid^{2} = v^{2} dT^{2} = c^{2} dT^{2}sin^{2}\theta,{\nobreakspace}and
\end{equation}
\begin{eqnarray}
\label{eq:11}
\mid\overline{AB}\mid = V_{T}^{2} dT^{2}=c^{2} dT^{2}cos^{2}\theta=\frac{c^{2} dT^{2}}{\gamma} = dW^{2}.
\end{eqnarray}

The Euclidean metric provides that the lengths of the sides of triangle
ABC, being a right triangle, obey the Pythagorean theorem:
\begin{equation}
\label{eq:12}
\mid\overline{BC}\mid^{2} = \mid\overline{AC}\mid^{2} + \mid\overline{AB}\mid^{2}{\nobreakspace}.
\end{equation}
To confirm that the Euclidean metric is obeyed, Equations{\nobreakspace}(\ref{eq:9})-(\ref{eq:11}) are substituted into Equation{\nobreakspace}(\ref{eq:12}). This gives:
\begin{eqnarray}
\label{eq:13}
c^{2} dT ^{2} = c^{2} dT^{2}sin^{2}\theta + c^{2} dT ^{2}cos^{2}\theta \nonumber\\
= c^{2} dT^{2}(sin^{2}\theta + cos^{2}\theta) = c^{2} dT ^{2} ,
\end{eqnarray}
which confirms self-consistency of the derivation. Meanwhile, the Lorentz metric gives the equation:
\begin{equation}
\label{eq:14}
ds^{2} = -c^{2}d\tau^{2} = - c^{2}dt^{2} + dx^{2}+dy^{2}+dz^{2} ,
\end{equation}
where \ensuremath{d\tau} is the proper time, \ensuremath{d\tau = dt/\gamma}. For a particle in motion, Equation{\nobreakspace}(\ref{eq:14}) becomes
\begin{equation}
\label{eq:15}
-c^{2}dt^{2}/\gamma^{2} = - c^{2}dt^{2} + v^{2}dt^{2} \tab .
\end{equation}
Comparing Equation{\nobreakspace}(\ref{eq:15}) to the right triangle \ensuremath{\overline{ABC}} in Figure{\nobreakspace}3, it is clear that Equation{\nobreakspace}(\ref{eq:15}) is Equation{\nobreakspace}(\ref{eq:12}) rearranged as:
\begin{equation}
\label{eq:16}
-\mid\overline{AB}\mid^{2} = -\mid\overline{BC}\mid^{2} + \mid\overline{AC}\mid^{2}\tab ,
\end{equation}
Thus, in this theory the Lorentz metric is the Euclidean metric rearranged to provide a statement specifying the rate at which events move along the \ensuremath{W} coordinate as a function of the velocity of particles.
The Lorentz metric includes time in its calculation by combining \ensuremath{t}, \ensuremath{T}, and \ensuremath{W} into one coordinate by defining fixed relationships among these independent variables. In particular it assumes that local time is global time \ensuremath{(dt\equiv dT)}, and it defines the relationship between \ensuremath{t} and \ensuremath{W} as:
\begin{equation}
\label{eq:17}
d\tau^{2} \equiv  dW^{2}/ c^{2}\tab  .
\end{equation}

\section{New Insights}

This section uses the new coordinate system to infer effects beyond the three dimensions that we observe. 
Correlations across neighboring observable universes are explored. Also, the observation of only three of 
four spatial dimensions is explored. Curvature is added to \ensuremath{F_{i}} to explore the expansion 
and closure of the universe. 
The principle of relativity is applied across time to determine  \ensuremath{dt/dT} as a 
function of space and time. From that, two ages of the universe are computed and compared, 
one from \ensuremath{t} and the other from \ensuremath{T}.

\subsection{Flow of Past, Present, Future}

For a flat section of the observable universe, if the vast preponderance of hyper-particles have 
non-relativistic observed velocity (\ensuremath{\gamma \approx 1} ) and travel towards
\ensuremath{-W (V_{W} \approx -c)}, then, from Equation{\nobreakspace}(\ref{eq:17}), any observable universe in the 
\ensuremath{W > 0} portion of space has a set of events (particle locations) that will 
later recur nearly identically at the observable universe at \ensuremath{W=0}. 
For example, the spatial configuration at \ensuremath{W = D{\nobreakspace}(D>0)} at global
time \ensuremath{T} will recur at \ensuremath{W=0} at time \ensuremath{T'=T+D/c}. Thus, defining the
observable universe at \ensuremath{W=0} as the "present", then the "future"
is at \ensuremath{W > 0}. Further, the "past" is at \ensuremath{W<0}. Configurations flow from \ensuremath{(W>0)} to  \ensuremath{(W=0)} to \ensuremath{(W<0)}. The farther one proceeds along the
\ensuremath{+W} direction, the farther into the "future" one goes, and the
farther one proceeds along \ensuremath{-W}, the farther into the "past" one
goes; however the correlation decreases because the difference
in flow rates along \ensuremath{W} (related to differences in observed velocity)
becomes increasingly important.

While \ensuremath{W} is correlated with a time coordinate, it is not a time
coordinate. Traveling in \ensuremath{W}, if possible, is not traveling in
time. It is traveling to a place whose spatial configuration is correlated 
with a spatial configuration that existed elsewhere
at another time, and the correlation is reduced, according to
Equation{\nobreakspace}(\ref{eq:17}), by the presence of relativistic objects whose
events travel through W at reduced speed.

\subsection{Slicing Hyperspace Into Parallel Universes}

The Lorentz formulation of relativity provides that for light
\ensuremath{d\tau =0}, which defines a light-cone. Equation{\nobreakspace}(\ref{eq:17})
provides a translation of that into the Euclidean formulation
as \ensuremath{dW=0}. This infers that hyper-particles of light do not travel
in the direction perpendicular to \ensuremath{F_{i}}. Thus, photons, and all light-like
particles, spend their lifetime entirely in the one observable
universe in which they were generated. Thus, each observable
universe is physically a sub-space containing all light-like particles that might
ever interact with one another.

Equation{\nobreakspace}(\ref{eq:4}), for light-like particles, gives \ensuremath{cos\theta=0}.
Thus, their hyper-particles are everywhere locally parallel to
the local observable universe. This implies, further, that their
entire hyper-particle is observed. Thus, the particle is the
hyper-particle, which implies that the derivation that led to
Equations{\nobreakspace}(\ref{eq:6})-(\ref{eq:8}) is not valid for light-like particles. Instead,
the X component of the hyper-particle speed must match the observed
speed, which implies
\begin{equation}
\label{eq:18}
V_{X} =v = \pm c \tab .
\end{equation}
Separately, for the W component of the hyper-particle motion,
dW=0, implies
\begin{equation}
\label{eq:19}
V_{W}=0 \tab ,
\end{equation}
i.e., photons have no velocity perpendicular to F. Thus,
Figure{\nobreakspace}3 and the Lorentz metric are valid for all particles,
including light-like particles.

\subsection{The Nature of the Observer}

Our observation of only the spatial dimensions \ensuremath{(x,y,z)} while residing in a 
Euclidean \ensuremath{(W,X,Y,Z)} space is data that provides new understanding of our 
nature as observers.  In the construction, \ensuremath{F_{0}} was defined as an observable universe, 
such that an observer in \ensuremath{F_{0}} observes only that one observable universe. 
Equation{\nobreakspace}(\ref{eq:19}) indicates that light-like hyper-particles in any \ensuremath{F_{i}} are trapped in 
that \ensuremath{F_{i}} and observe only that which is in that observable universe. That is, 
light-like particles, are limited to observing \ensuremath{(x,y,z)}. 
All other hyper-particles experience many observable universes.
Thus, some key aspect of us as observers, e.g. the mechanism of memory and/or
observation, though not necessarily the physical body, must be fundamentally light-like. 

Each observer sees past, present, and future unfold in one observable universe. The spatially 
correlated configurations on other observable universes are not something that we typically see. 
Those configurations are simply correlated configurations, and by the time they come to 
pass on the observable universe where a particular observer sees them, those configurations no
longer exist on any other observable universe.

For each observer, the calculation of Lorentz distances is correct
over a region of space in which the path of light is perpendicular
to the hyper-particle of the observer. For example, in Figure
1, an observer at Particle 1 can correctly assume a flat observable
universe in calculating the Lorentz distances between it and
Particles 2 and 3. Observers at Particles 2 and 3 cannot, however,
correctly assume that their flat observable universe extends
to either of the other particles. In trying to do so, observers
at Particles 2 and 3 obtain the wrong answers. This resolves the
twin paradox\cite{Gravitation} in the Euclidean formulation.

The observer at Particle 1 in Figure 1 is special because it
is at rest with respect to the larger observable universe. It
also has the fastest rate of aging. The shape of the observable
universe can be mapped by measuring the velocity at which the
clock rate is maximal at various locations. This does
not imply a global rest frame, since there is no mechanism that
fixes the motion of any observable universe relative to the Euclidean
frame. Also, it does not violate relativity since the local observer
at Particle 1 cannot determine that it is special by any local
measurement.

\subsection{Kinetic Energy Is Mass Energy}

Energy conservation is required within each observable universe
for compliance with both Newtonian and relativistic mechanics.
Since hyper-particles flow through the observable universe with
a fixed speed \ensuremath{\mid V_{T}\mid=c}, by Equation{\nobreakspace}(\ref{eq:8}), hence, if hyper-particles
have mass energy per unit length, then that mass energy density
is constant over the length of the hyper-particle. Otherwise
the quantity of mass energy entering the sub-space would be different
from the quantity leaving the sub-space, and energy would not
be conserved in the sub-space.

The mass energy of the observed particle is the product of the
mass energy per unit length, \ensuremath{\rho_{m}c^{2}}, times the length of
hyper-particle within the observed universe, \ensuremath{\epsilon '= \epsilon/cos\theta}.
Defining \ensuremath{m=\rho_{m} \epsilon} as the rest mass of the particle, then the mass energy 
of a particle at any velocity is
\begin{equation}
\label{eq:20}
E= \gamma m c^{2} \tab . 
\end{equation}
The mass energy in Equation{\nobreakspace}(\ref{eq:20}) corresponds with the combined value for the
kinetic energy plus mass energy in special relativity. Hence,
in the Euclidean formulation, the kinematic energy of the particle
is a contribution to the mass energy. Observed particles have no kinetic energy apart from their
mass energy.

As the velocity of an observed particle changes, the length of
the corresponding hyper-particle within the observable universe
changes. Yet, since \ensuremath{V_{T}} is a constant, hence the 
rate of hyper-particle entering and leaving the observable universe remains unchanged.
Hence, the energy required to generate increased length comes
from inside the observable universe, perhaps exchanged between
hyper-particles within the observable universe by forces within
that observable universe.

Special relativity provides momentum as \ensuremath{\overline{p}=\gamma m \overline{v}}.
Since each of the terms on the right of this equation is defined
in the new formulation in accord with special relativity, the
mathematical quantity \ensuremath{\overline{p}} can be defined in Euclidean \ensuremath{(W,X,Y,Z)}
space as an observable three-vector locally within any observable
universe. Since the energy, \ensuremath{E}, is also in accord with special
relativity, it follows that, mathematically, the equations
for conservation of this momentum are identical to the Lorentz formulation. 

In this formulation the interpretation of momentum changes slightly because the mass 
of the particle is physically a function of its velocity. That is,  \ensuremath{\gamma m} is
physically the mass of the hyper-particle. The interpretation appears to apply even for photons 
traveling at the speed of light. For photons, the entire hyper-particle is observed. Thus, 
their energy is the mass of the
entire hyper-particle and the momentum is that mass times the velocity, \ensuremath{c}.

\subsection{Concentric Spherical Universes}

\label{sec:sphere}

The ubiquitous redshift of light from distant astronomic phenomena
may be interpreted as a result of the expansion of the universe\cite{astrophys}.
The Hubble Constant, \ensuremath{H}, indicates,
in particular, a linear relation (at least approximately) between the distance to an object and the
size of the red shift. One approach to modeling such expansion
is to give the observable universe a curvature of radius, \ensuremath{R},
with an expansion rate \ensuremath{dR/dT}. If the relation between distance and red shift
is linear, then R is constant, otherwise R may be a function of position.

With the addition of this curvature, within the Euclidean formulation,
each observable universe becomes a hollow sphere (if R is constant) whose surface
area and volume is finite and which grows over time. The thickness, \ensuremath{\epsilon},
also must increase over time in proportion, by relativity. 

Figure 4 shows a slice through a set of concentric spherical
observable universes. It highlights three select observable universes
of a continuous set. At every location there is a local \ensuremath{w} coordinate
that corresponds by Euclidean transformation (spatial rotation
and translation) to a global radial coordinate, \ensuremath{R}. From a global
perspective, the observable universes are seen to expand outward
towards \ensuremath{+R}, while locally hyper-particles
flow towards \ensuremath{-w} at the local speed of light. Figures 1 through 3, may
be viewed as close-ups of a very small and therefore nearly flat
sections of the curved observable universes shown in Figure 4.

\begin{figure*}
\includegraphics[width=7in]{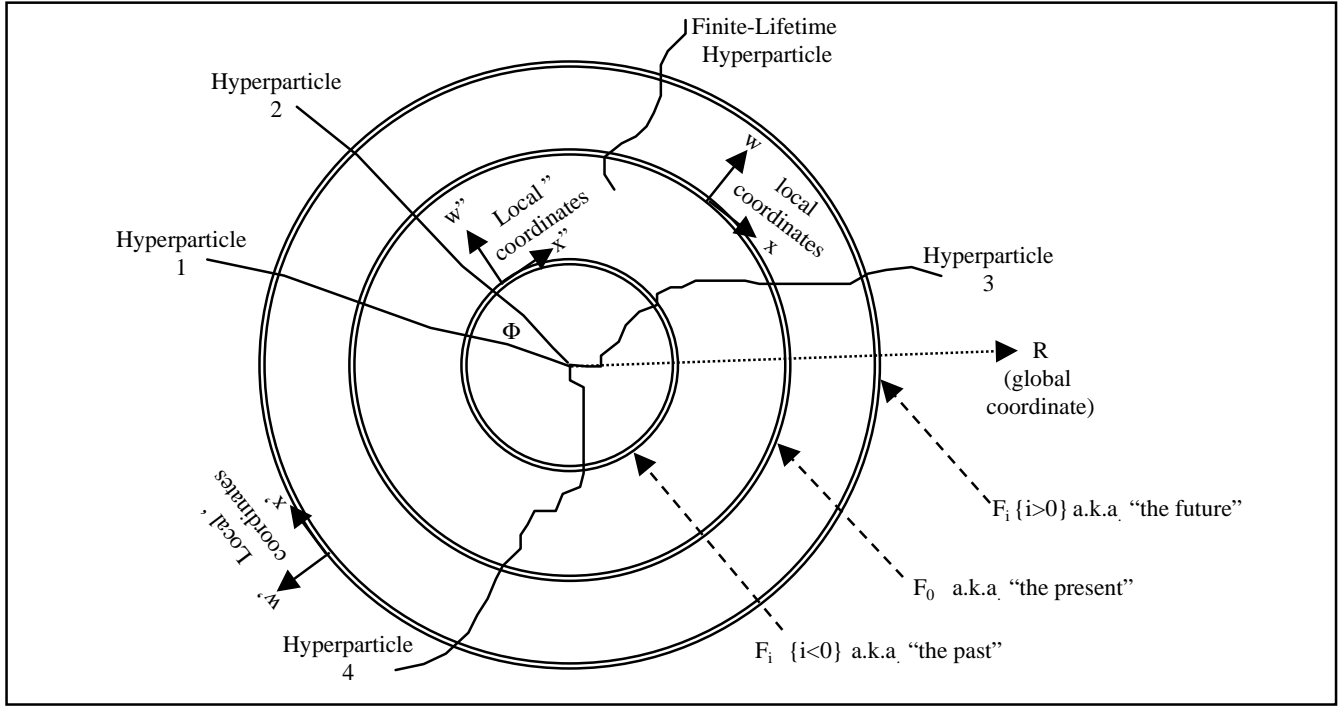}
\caption{The (W,X,Y,Z) Euclidean universe as a series of
expanding concentric (x,y,z) spherical observable universes correlated
as ``past'', ``present'' and ``future''. Each sphere expands outward. By relativity
no observer can tell, by local measurement which observable universe it is on.}
\end{figure*}

Extending the principle of relativity across observable universes,
it is possible to determine many properties of the universe expansion.
Of course, since the corpus of experimental results all apply
to one observable universe, there is no direct measurement that
indicates that relativity applies across observable universes.
However, that aside, relativity implies the following constraints
in order that an observer cannot distinguish observable universes:
\begin{enumerate}
\item \ensuremath{c \propto R} so that the circumference of each observable universe
is the same number of wavelengths for all \ensuremath{R};
\item \ensuremath{\epsilon \propto R} so that the ratio of the circumference
to the thickness is the same for each observable universe; 
\item The rate of aging is independent of \ensuremath{R}, by Equation{\nobreakspace}(\ref{eq:1}) combined
with Constraints 1 and 2 above.
\end{enumerate}

\subsection{Two Ages for Our Universe}

An observable universe has two ages\cite{9905022}:
\begin{itemize}
\item Time in current seconds since the start of the universe as computed
from the Hubble Constant, \ensuremath{H}
\item Evolutionary age computed as the integral of \ensuremath{dt} from \ensuremath{T}=0 to the
present time.
\end{itemize}

The value of \ensuremath{H} is determined by measurement of the expansion-induced
relative velocity between particles some distance apart in the
observable universe. That velocity is observed not with a ruler,
but rather as a change in the time for light to travel between
two particles. For Hyper-particles 1 and 2 in Figure 4, the observed
velocity of expansion, in global time units, is
\begin{equation}
\label{eq:21}
V_{exp} =c\frac{d(\Phi R/c)}{dT}\tab  .
\end{equation}
where \ensuremath{\Phi} is a fixed angle between the particles as
shown in Figure{\nobreakspace}4. This may be rewritten as
\begin{equation}
\label{eq:22}
V_{exp} =c(\frac{\partial (\Phi R/c)}{\partial R} \frac{dR}{dT}+ 
\frac{\partial (\Phi R/c)}{\partial T} )
\tab .
\end{equation}

Since \ensuremath{c \propto R}, hence \ensuremath{\Phi R/c} is independent of \ensuremath{R} so
the first term is zero. Thus,
\begin{equation}
\label{eq:23}
V_{exp} = - \frac{\Phi R}{c} \frac{\partial c}{\partial T} \tab .
\end{equation}

By relativity, to prevent the distinction of observable universes,
\ensuremath{V_{exp}} must be proportional to \ensuremath{c}, so that their ratio cannot be
used to determine location in space. Using this in Equation{\nobreakspace}(\ref{eq:23})
gives a first order non-linear differential equation:
\begin{equation}
\label{eq:24}
c \propto - \frac{\Phi R}{c} \frac{\partial c}{\partial T} \tab ,
\end{equation}
which has the solution
\begin{equation}
c \propto R/T \tab .
\label{eq:cproptoRT}
\end{equation}
Equation{\nobreakspace}(\ref{eq:cproptoRT}) can be recast in terms of the current speed
of light as
\begin{equation}
\label{eq:26}
c=c_{0}(R/R_{0}) (T_{0}/T)\tab ,
\end{equation}
where \ensuremath{T_{0}} is the present global time, \ensuremath{c_{0}} is the present speed of light at \ensuremath{F_{0}}, and \ensuremath{R_{0}} is the present radius of \ensuremath{F_{0}}.
Thus, \ensuremath{c} at any location decreases over time. Also, substituting
Equation{\nobreakspace}(\ref{eq:26}) into Equation{\nobreakspace}(\ref{eq:21}), gives
\begin{equation}
V_{exp} =\frac{\Phi R}{T} \tab  .
\label{eq:vexp}
\end{equation}
Since \ensuremath{H} is the ratio of the velocity between particles and the
distance between particles, hence
\begin{equation}
H(T)= \frac {V_{exp}}{\Phi R} = \frac{1}{T}\tab ,
\end{equation}
which, finally gives the age of the universe in global seconds as
\begin{equation}
T = 1/H   \tab .
\end{equation}
Meanwhile, the rate of local time, obtained by using Equation{\nobreakspace}(\ref{eq:26}) in Equation{\nobreakspace}(\ref{eq:1}) is
\begin{equation}
\frac{dt}{dT}= \frac{T_{0}} {T} \tab ,
\label{eq:dtdT}
\end{equation}
which integrates to give the evolutionary age of the universe as
\begin{eqnarray}
\int _{0}^{T_{0}}\!\!\! dt=  \int _{0}^{T_{0}}\!\!\! dT \frac{T_{0}}{T} 
= T_{0}log(\frac{T_{0}}{T}) \mid_{0}^{T_{0}}\nonumber\\
=   -T_{0}log(0)=\infty{\nobreakspace}.
\label{eq:infiniteage}
\end{eqnarray}
Thus, the total aging since the beginning of global time is predicted
to be infinite everywhere in the universe even though the age
of the universe, in global time units, is finite. This prediction
is consistent with data that has generally shown, until recently reinterpreted to reduce the
discrepancy \cite{freedman2}, that the age of
the universe, as obtained from observations of astronomic evolution,
is greater than that computed from the Hubble Constant.

\subsection{Refraction Closing Our Universe Without Gravity or Dark Matter}

The spherical shape for each observable universe, \ensuremath{F_{i}}, means that its extent in each 
local dimension \ensuremath{x,y,z} is finite and wraps around, i.e., if one goes far enough 
along \ensuremath{+x} one ends up a \ensuremath{-x}. The volume for each observable universe is finite. 
Meanwhile the volume
for the global \ensuremath{(W,X,Y,Z)} universe is infinite and there are an infinite
number of observable universes.

The spherical shape of \ensuremath{F_{i}} is consistent with refraction-induced curvature. Applying Huygen's principle to the \ensuremath{R}-dependent speed of light as given in Equation{\nobreakspace}(\ref{eq:26}), a wavefront of light starting at Hyper-particle 1 (in Figure{\nobreakspace}4), for example, propagates an equal number of wavelengths
at each radius. Because \ensuremath{c \propto R}, hence the wavefront proceeds
as a straight line rotating around \ensuremath{R=0}. Thus, by refraction each
point on the wavefront travels in a circle.

This refraction does not appear to be a gravitational lensing effect.\cite{gravity} There are two reasons. 
First, gravity is a force associated with mass. In particular, gravity is a function of the distance from the mass (measured along a curve within \ensuremath{F_{i}}). Thus, while the three-dimensional deformation of \ensuremath{F_{i}} near each hyper-particle may be gravity-induced refraction (see Section \ref{sec:futuredirections}), the \ensuremath{R}-dependent gradient of the speed of light has the wrong spatial dependence to be related to the distance from mass.

Second, since gravity is a force, hence, if it is present, then there should be a tell-tale deceleration
of \ensuremath{F_{i}}. In particular, the gravitational force on the observable universe, \ensuremath{F_{0}} at \ensuremath{R_{0}} would be proportional
to the integral of the mass from \ensuremath{R=0} to \ensuremath{R= R_{0}}. Instead, by relativity, in order that observable universes cannot be distinguished by the ratio of the expansion rate to the speed of light, and by Equation{\nobreakspace}(\ref{eq:cproptoRT}), hence the rate of radial motion at any point in the universe at any time must be:
\begin{equation}
V_{R} \propto  c \propto R/T{\nobreakspace}.
\label{eq:VR}
\end{equation}
Equation (\ref{eq:VR}) can be manipulated to compute the time dependence of the radius, \ensuremath{R_{i}(T)}, of any particular observable universe, \ensuremath{F_{i}}. First, representing \ensuremath{V_{R}} as a differential gives:
\begin{equation}
\frac{dR}{dT} \propto R/T \tab,
\end{equation}
which can be rearranged as:
\begin{equation}
\frac{dR}{R} \propto \frac{dT}{T} \tab.
\end{equation}
Both sides are integrated to give
\begin{equation}
(R_{i}(T) - R_{i}(0))\propto T\tab ,
\label{eq:Radiusvstime}
\end{equation}
where \ensuremath{R_{i}(0)} is the radius of \ensuremath{F_{i}} at \ensuremath{T=0}.

Equation{\nobreakspace}(\ref{eq:Radiusvstime}) shows that the rate of expansion for any particular observable universe, in global time units, is constant - there is no deceleration. Hence, either the gravitational force is cancelled by another force, or it does not exist on this grand scale or in the \ensuremath{R} direction. In any case, the refraction that closes each observable universe exists by some cause other than gravity.

\subsection{A Continuous, Accelerating Universe}

Equations{\nobreakspace}(\ref{eq:cproptoRT}) and (\ref{eq:vexp}) show singularities occuring at \ensuremath{T=0}, similar to those implied by the "Big Bang" theories. However, Equation{\nobreakspace}(\ref{eq:infiniteage}) describing the evolutionary age of the universe applies for all times \ensuremath{T}. Thus, even when the universe was one global second old, the evolutionary age was already infinite. 

In the early universe, clocks ran faster so the rate of system evolution was higher. The general relationship between the global and local rates of time can be derived by changing the integration limits on Equation{\nobreakspace}(\ref{eq:infiniteage}). This gives:
\begin{equation}
T = T_{0}e^{(t-T_{0})/T_{0}}\nobreakspace.
\label{eq:Tvst}
\end{equation}
With this normalization, \ensuremath{t} and \ensuremath{T} have the same values and derivatives at time \ensuremath{T_{0}}. When \ensuremath{t=0}, \ensuremath{T = T_{0}e^{-1}}, and when \ensuremath{t=-\infty}, \ensuremath{T=0}. When the universe was one global second old, the local clock rate on our observable universe, \ensuremath{F_{0}}, would have been approximately \ensuremath{5x10^{17}} times its current rate (using 16 Billion years as its current age). 

The rapid clock rate of the early universe may have been undetectable by local measurement. For example, the circumnavigation time in local seconds, \ensuremath{\Delta t_{c}}, for light to complete a trip around a particular observable universe was the same when the universe was one second old as it is now. The circumnavigation time is obtained by combining Equations (\ref{eq:cproptoRT}), (\ref{eq:dtdT}), and (\ref{eq:Radiusvstime}). This gives:
\begin{equation}
\Delta t_{c} \propto \frac{T_{0}}{T} \Delta T_{c}
\propto \frac{T_{0}}{T} \frac{2\pi R_{i}(T)}{c_{i}(T)}
\propto \frac{R_{0}(T_{0})}{c_{0}(T_{0})}\nobreakspace,
\end{equation}
where \ensuremath{c_{i}(T)} is the speed of light vs. time on observable universe \ensuremath{F_{i}}, and \ensuremath{\Delta T_{c}} is the circumnagivation time in global seconds. Thus, the observed circumnagivation time is the same on all observable universes at all times, even when the universe was one global second old.

Some of the predicted phenomona of the "Big Bang" theories may still apply. For instance, if \ensuremath{R_{i}(0)=0}, then all of the observable universes were collapsed to a point at T=0. Conversely, if \ensuremath{R_{i}(0)>0}, then the observable universes did not collapse to a point at T=0.

Our interpretation of the expansion rate of the universe is also affected by local time. Equation{\nobreakspace}(\ref{eq:Radiusvstime}) shows a constant rate of expansion with respect to global time.  However, substituting Equation{\nobreakspace}(\ref{eq:Tvst}) into Equation{\nobreakspace}(\ref{eq:Radiusvstime}) gives the expansion as a function of local time:
\begin{equation}
(R_{i}(T) - R_{i}(0))\propto T_{0}e^{(t-T_{0})/T_{0}}\nobreakspace.
\label{eq:Radiusvslocal}
\end{equation}
Thus, the universe is expanding exponentially with respect to local time, i.e., the expansion is accelerating\cite{9906220}\cite{accell2}\cite{Repulsion}.

\subsection{A New Lagrangian}

The motion of \ensuremath{F_{i}} even absent the presence of hyper-particles indicates that space has substance as a continuum in accord with other models\cite{9905066}\cite{scalar-tensor}. A Lagrangian density, \ensuremath{L}, is defined here to describe that continuum as an ideal cosmological fluid:
\begin{equation}
L(X_{\mu}) =\frac{\phi}{2} V_{\mu} V_{\mu} -\frac{k}{2}
\frac{\partial\phi}{\partial X_{\mu}}\frac{\partial\phi}{\partial X_{\mu}}
+\lambda(\frac{\partial\phi}{\partial T} +
 \frac{\partial(\phi V_{\mu})} {\partial X_{\mu}})
{\nobreakspace}.
\label{eq:lagrangian}
\end{equation}
Here \ensuremath{X_{\mu}} represents \ensuremath{\{W,X,Y,Z\}}, \ensuremath{\phi(W,X,Y,Z,T)} is the time-dependent density of the cosmological fluid, \ensuremath{V_{\mu}(W,X,Y,Z,T)} is the time-dependent velocity of the fluid, \ensuremath{k} is a real constant of currently unknown value, and \ensuremath{\lambda}(W,X,Y,Z,T) is a Lagrange multiplier\cite{variationalprinciples}.  The first term of the Lagrangian density is the kinetic energy density; the second term is a harmonic potential energy density; and the third term is continuity as a holonomic constraint. Because the fluid is an 'ideal' fluid, there are no terms in the Lagrangian density for internal properties of the fluid such as temperature or entropy. The holonomic constraint requires that fluid cannot simply disappear in one place and reappear in another; it requires that transport of fluid occur via its motion from location to location. 

The Lagrangian is the four dimensional integral of the Lagrangian density over the spatial coordinates \ensuremath{(W,X,Y,Z)}. The Lagrangian gives rise to two Lagrange equations, a vector equation from the variation of the Lagrangian with respect to \ensuremath{V_{\mu}} and a scalar equation from its variation with respect to \ensuremath{\phi}. The vector Lagrange equation, for any Lagrangian density that is a function of \ensuremath{V_{\mu}} and its first derivatives but not higher order derivatives, is given by:
\begin{equation}
\frac{\partial L}{\partial V_{\mu}} 
-\frac{\partial}{\partial T} 
\frac{\partial L}{\partial \frac{\partial V_{\mu}}{\partial T}}
-\frac{\partial}{\partial X_{\mu}} \frac{\partial L}{\partial \frac{\partial V_{\mu}}{\partial X_{\mu}}}
=0 \nobreakspace.
\end{equation}
Substituting in \ensuremath{L} from Equation{\nobreakspace}(\ref{eq:lagrangian}), this vector equation reduces to:
\begin{equation}
V_{\mu}=\partial\lambda/\partial X_{\mu}  \tab.
\label{eq:lagrangianvmu}
\end{equation}
This result is general to all solutions of this Lagrangian and reduces the theory to a bi-scalar theory, where \ensuremath{\phi} and \ensuremath{\lambda} are the two scalar fields.

The scalar Lagrange equation, for any Lagrangian density that is a function of \ensuremath{\phi} and its first derivatives but not higher order derivatives, is given by:
\begin{equation}
\frac{\partial L}{\partial \phi} 
-\frac{\partial}{\partial T} \frac{\partial L}{\partial \frac{\partial \phi}{\partial T}}
-\frac{\partial}{\partial X_{\mu}} \frac{\partial L}{\partial \frac{\partial \phi}{\partial X_{\mu}}}
=0 \nobreakspace .
\end{equation}
Substituting \ensuremath{L} from Equation{\nobreakspace} (\ref{eq:lagrangian}) and reducing by use of Equation{\nobreakspace}(\ref{eq:lagrangianvmu}) gives
\begin{equation} 
k\frac{\partial^{2}\phi}{\partial X_{\mu}^{2}}
- \frac{1}{2}\frac{\partial \lambda}{\partial X_{\mu}}\frac{\partial \lambda}{\partial X_{\mu}}
- \frac{\partial \lambda}{\partial T} =0{\nobreakspace},
\end{equation}
which which must be solved together with the continuity equation. The continuity equation, after substitution from Equation{\nobreakspace}\ref{eq:lagrangianvmu}, is
\begin{equation}
\frac{\partial \phi}{\partial T} + 
\frac{\partial} {\partial X_{\mu}}( \phi \frac{\partial \lambda}{\partial X_{\mu}}) = 0{\nobreakspace}.
\end{equation}
A closed form solution to the Lagrange equation that satisfies continuity, is:
\begin{equation}
\phi = \frac{K}{T^{N}} 
{\nobreakspace\nobreakspace}and{\nobreakspace\nobreakspace} 
\lambda = 1/2 \frac{R^{2}}{T} \nobreakspace, 
\label{eq:lagrangiansolution}
\end{equation}
where \ensuremath{N} is the number of spatial dimensions of the universe [which is at least four to include \ensuremath{(W,X,Y,Z)}], and where \ensuremath{R=(X_{\mu}X_{\mu})^{1/2}}. The value of \ensuremath{K} is undetermined. At \ensuremath{T=0} both scalar fields are infinite exposing the same "Big Bang" type singularity as seen earlier. 

The dimensional dependence of \ensuremath{\phi} in Equation{\nobreakspace} (\ref{eq:lagrangiansolution}) concurs with the analysis in Section{\nobreakspace}3.5. In particular, it shows the volume of each observable universe expanding as \ensuremath{T^{4}} for a \ensuremath{(W,X,Y,Z)} space, which corresponds to each observable universe expanding both due to the increasing radius and due to its increasing thickness, \ensuremath{\epsilon}, as predicted by relativity arguments in the earlier section.

The radial velocity of each point in the universe is computed by differentiating \ensuremath{\lambda} as per Equation{\nobreakspace}(\ref{eq:lagrangianvmu}). The result is
\begin{equation}
\ensuremath{V_{R} = R/T}\nobreakspace,
\label{eq:vrsolution} 
\end{equation}
which is consistent with Equation (\ref{eq:VR}) and determines that \ensuremath{R_{i}(0)=0}.  Thus, for this
solution, the universe was collapsed to a point at \ensuremath{T=0} which would be in accord with "Big Bang" theories.

An alternate closed form solution for this Lagrangian is:
\begin{eqnarray}
\phi = \frac{-(N+1)}{(2k)(N+2)(4+2N)^{2}}\frac{R^{4}}{T^{2}}\nobreakspace, 
{\nobreakspace}and{\nobreakspace} \\
\lambda = \frac{1}{4+2N} \frac{R^{2}}{T}\tab\tab\nobreakspace. 
\label{eq:lagrangiansolution2}
\end{eqnarray}
This solution provides \ensuremath{R_{i}(0)>0}. Thus if the actual universe obeys this solution, then the  universe was not collapsed to a point at \ensuremath{T=0}, contrary to "Big Bang" theories.

Both of the solutions show closure of the universe even before hyper-particles or mass are introduced.

\section{Future Directions}
\label{sec:futuredirections}

A key future goal is to incorporate hyper-particles into the Lagrangian as perturbations on \ensuremath{\phi} and \ensuremath{\lambda}, and to solve thereby for the structure of the fundamental particles. That further may lead to a derivation of the forces between particles as interactions among those perturbations and to a determination of the fundamental constants and any scaling of those constants over time and space. It may further lead to an exploration of a four-dimensional Euclidean momentum with possible conservation laws and, separately to an exploration of the relation between the presented theory and standard string theories. 

Mapping general relativity into the Euclidean formulation may be achievable by an investigation of the relationship between gravity and the deformations of \ensuremath{F_{i}} near hyper-particles. The possibility of non-gravitationally induced refraction on astronomic scales may imply that some phenomena currently ascribed to dark matter \cite{darkmatter} may actually be unrelated to gravity and due rather to the new phenomenon shown here to cause the refraction that closes the universe.

Other unknowns to be investigated include the possibility of freeing oneself from a particular observable universe to directly observe others, and the possibility of additional Euclidean dimensions as might be inferred from experimental observations.  Additionally, a goal is to determine the thickness of the observable universe (a value for \ensuremath{\epsilon}). Also the ratio between the expansion rate of the universe and the speed of light is of fundamental interest.

\section{Conclusions}

The separation of Lorentz time into three parameters -- global time, a global spatial coordinate, and a local rate of aging that is a function of position and time, -- has enabled successful formulation of relativity in a global \ensuremath{(W,X,Y,Z,T)} Euclidean space.  The new formulation agrees with the main predictions of special relativity, provides re-interpretation of some relativistic phenomena, and provides new predictions. Some of the predictions are supported by existing experimental evidence. Others may be tested by future experiments. The predictions are derived by the application of the principle of relativity via geometry and supported by a new Lagrangian describing the universe as an ideal cosmological fluid. Sample predictions include:

\begin{itemize}
\item The age of the universe as measured by the Hubble constant is
finite, while that measured by system evolution is infinite.
\item The universe is closed by refraction without gravity or dark matter, implying a new phenomenon at astronomic scales.
\item The speed of light varies throughout the universe.
\item Parallel universes exist and are correlated as {\textquotedbl}past{\textquotedbl}, {\textquotedbl}present{\textquotedbl}, {\textquotedbl}future{\textquotedbl}.
\end{itemize}

\end{document}